\documentclass[12pt]{article}

\usepackage{paralist}
\usepackage{amsfonts}

\usepackage{amsmath}
\usepackage{amssymb}
\usepackage{amsfonts}

\begin{document}

\newtheorem{theorem}{Theorem}
\newtheorem{corollary}{Corollary}

\fontsize{12}{6mm}\selectfont
\setlength{\baselineskip}{2em}

$~$\\[.35in]
\newcommand{\dss}{\displaystyle}
\newcommand{\raro}{\rightarrow}
\newcommand{\be}{\begin{equation}}

\def\sech{\mbox{\rm sech}}
\def\sn{\mbox{\rm sn}}
\def\dn{\mbox{\rm dn}}
\thispagestyle{empty}

\begin{center}
{\Large\bf A Geometric Method to Investigate Prolongation }  \\    [2mm]
{\Large\bf Structures for Differential Systems With Applications to }  \\    [2mm]
{\Large\bf Integrable Systems}  \\
\end{center}

\vspace{1cm}
\begin{center}
{\bf Paul Bracken}                        \\
{\bf Department of Mathematics,} \\
{\bf University of Texas,} \\
{\bf Edinburg, TX  }  \\
{78541-2999}
\end{center}

\vspace{3cm}
\begin{abstract}
A type of prolongation structure for several general systems is discussed.
They are based on a set of one-forms in which the underlying structure 
group of the  integrability condition
corresponds to the Lie algebra of $SL (2, \mathbb R)$,
$O(3)$ or $SU (3)$. Each will be considered in turn and 
the latter two systems represent larger $3 \times 3$ cases.
This geometric  approach is
applied to all three of these systems to obtain 
prolongation structures explicitly. In both $3 \times 3$ cases, the prolongation 
structure is reduced to the situation
of three smaller $2 \times 2$ problems. Many types of conservation
laws can be obtained at different stages of the development, and at the end,
a single result is developed to show how this can be done.
\end{abstract}

\vspace{2mm}

\vspace{2mm}
MSCs: 35A30, 32A25, 35C05

\newpage
\begin{center}
{\bf I. Introduction.}
\end{center}

Geometric approaches have been found useful in producing a great
variety of results for nonlinear partial differential equations {\bf [1]}. 
A specific geometric approach discussed here
has been found to produce a very elegant, coherent
and unified understanding of many ideas in nonlinear physics
by means of fundamental differential geometric concepts.
In fact, relationships between a geometric interpretation of
soliton equations, prolongation structure, Lax pairs and
conservation laws can be clearly realized and made use of. 
The interest in the approach, its generality and the results
it produces do not depend on a specific equation at the outset. The  
formalism in terms of differential forms {\bf [2]} can encompass large classes of
nonlinear partial differential equation, certainly the AKNS systems {\bf [3,4]}, 
and it allows the production of generic expressions for
infinite numbers of conservation laws. Moreover, it leads to the
consequence that many seemingly different equations turn out to be
related by a gauge transformation.

Here the discussion begins by studying prolongation structures
for a $ 2 \times 2$ $SL (2, \mathbb R)$ system discussed first by 
Sasaki {\bf [5,6]} and Crampin {\bf [7]} to present and illustrate
the method. This will also demonstrate the procedure and the kind
of prolongation results that emerge. It also provides a basis from which
to work out larger systems since they can generally be reduced to 
$2 \times 2$ problems. Of greater complexity are a pair of 
$3 \times 3$ problems which will be considered next. In particular,
it is shown how to construct an $O (3)$ system based on three
constituent one-forms as well as an $SU (3)$ system composed of
eight fundamental one-forms. The former has not appeared and in both
cases, all of the results are presented explicitly. The approach is 
quite unified and so once the formalism is established for the
$SL ( 2, \mathbb R)$ system, the overall procedure can be carried over to
the other Lie algebra cases as well.

Of course, the $3 \times 3$ problems will yield more types of 
conservation laws. The prolongation structures of the $3 \times 3$
problems will be observed to be reduceable to the same type as the  
$2 \times 2$ system considered at the beginning. However, the Riccati 
representations become much more complicated {\bf [8]}.
The generalization of this formalism to an $n \times n$ problem
then becomes straightforward after completing the study of these
smaller cases. The prolongation structure of an $n \times n$ 
problem is reduced to a collection of $n$ smaller $(n-1) \times (n-1)$
cases, and finally in the end to a set of $2 \times 2$ problems
such as the $S L (2, \mathbb R)$ case at the beginning.
Finally, to summarize at the end, a collection of conservation
laws is developed from one of the results and some speculation 
as to how how this type relates to other procedures {\bf [9]}.

\begin{center}
{\bf 2. $SL (2, \mathbb R)$ Prolongations.}
\end{center}

The procedure begins by associating a system of Pfaffian equations
to the nonlinear system to be studied, namely,
\be
\alpha_i =0,
\qquad
\alpha_i = d y_i - \Omega_{ij} \, y_j,
\qquad
i,j=1,2.
\label{eq2.1}
\end{equation}
In \eqref{eq2.1}, $\Omega$ is a traceless $2 \times 2$ matrix
which consists of a family of one-forms, $\omega_i$. 
In the end, it is desired to express these one-forms
in terms of independent variables, which are called $x$ and $t$,
the dependent variables and their derivatives. However, no
specific equation need be assumed at the start. The underlying structure
group is $S L (2, \mathbb R)$. Explicitly, the matrix of one-forms 
is given by the $2 \times 2$ matrix
\be
\Omega =
\begin{pmatrix}
\omega_1  &  \omega_2  \\
\omega_3  &  - \omega_1  \\
\end{pmatrix}
\label{eq2.2}
\end{equation}
The integrability conditions are expressed as the vanishing of
a traceless $2 \times 2$ matrix of two-forms $\Theta$ given by
\be
\Theta = d \Omega - \Omega \wedge \Omega.
\label{eq2.3}
\end{equation}
In terms of components, the left-hand side of \eqref{eq2.3}
has the form
\be
\Theta = ( \Theta_{ij} ) =
\begin{pmatrix}
\vartheta_1  &   \vartheta_2  \\
\vartheta_3  & - \vartheta_1  \\
\end{pmatrix}.
\label{eq2.4}
\end{equation}
Substituting \eqref{eq2.2} into \eqref{eq2.3}, a matrix which
contains the integrability equations in terms of the basis 
one-forms is obtained. The three equations are
\be
\vartheta_1 = d \omega_1 - \omega_2 \wedge \omega_3,
\qquad
\vartheta_2 = d \omega_2 - 2 \omega_1 \wedge \omega_2,
\qquad
\vartheta_3 = d \omega_3 +2 \omega_1 \wedge \omega_3.
\label{eq2.5}
\end{equation}
Of use in the theorems which follow, it is also useful to
have the $d \omega_i$ in \eqref{eq2.5} expressed explicitly
in terms of the $\vartheta_i$ as
\be
d \omega_1 = \vartheta_1 + \omega_2 \wedge \omega_3,
\qquad
d \omega_2 = \vartheta_2 + 2 \omega_1 \wedge \omega_2,
\qquad
d \omega_3 = \vartheta_3 - 2 \omega_1 \wedge \omega_3.
\label{eq2.6}
\end{equation}
Therefore, by selecting a particular choice for the set
$\{ \omega_i \}$, the nonlinear equation of interest can be
written simply as
\be
\Theta =0,
\qquad
\vartheta_i =0,
\qquad
i=1,2,3.
\label{eq2.7}
\end{equation}
It is easy to see that the system is closed.
Upon differentiating $\Theta$ and substituing \eqref{eq2.3},
we obtain
\be
d \Theta = \Omega \wedge \Theta - \Theta \wedge \Omega.
\label{eq2.8}
\end{equation}
This implies that the exterior derivatives of the set
of $\{ \vartheta_i \}$ are contained in the ring of two-forms
$\{ \vartheta_i \}$.

Now the differential ideal can be prolonged by including
the forms $\{ \alpha_1, \alpha_2 \}$ given by \eqref{eq2.1}.

{\bf Theorem 2.1.} The differential one-forms
\be
\alpha_1 = d y_1 - \omega_1 y_1 - \omega_2 y_2,
\qquad
\alpha_2 = d y_2 - \omega_3 y_1 + \omega_1 y_2,
\label{eq2.9}
\end{equation}
have the following exterior derivatives,
\be
d \alpha_1 = \omega_1 \wedge \alpha_1 + \omega_2 \wedge \alpha_2
- y_1 \vartheta_1 - y_2 \vartheta_2,
\quad
d \alpha_2 =- \omega_1 \wedge \alpha_2 + \omega_3 \wedge \alpha_1
+ y_2 \vartheta_1 - y_1 \vartheta_3.
\label{eq2.10}
\end{equation}
Hence, the exterior derivatives of the $\{ \alpha_i \}$ are
contained in the ring of forms $\{ \vartheta_i, \alpha_j \}$.

{\bf Proof:} The forms \eqref{eq2.9} follow by substituting
\eqref{eq2.2} into \eqref{eq2.1}.
Differentiating $\alpha_1$ in \eqref{eq2.9}
it is found that
$$
d \alpha_1 =-y_1 \, d \omega_1 + \omega_1 \wedge d y_1 - y_2 d \omega_2 
+ \omega_2 \wedge d y_2.
$$
Obtaining $dy_1$ and $d y_2$ from \eqref{eq2.9} and $d \omega_1$,
$d \omega_2$ from \eqref{eq2.6}, $d \alpha_1$ becomes,
$$
d \alpha_1 =- y_1 \vartheta_1 - y_1 \omega_2 \wedge \omega_3
+ \omega_1 \wedge \alpha_1 + y_2 \omega_1 \wedge \omega_2 -
y_2 \vartheta_2 - 2 y_2 \omega_1 \wedge \omega_2 + 
\omega_2 \wedge \alpha_2 + y_1 \omega_2 \wedge \omega_3 - y_2
\omega_2 \wedge \omega_1
$$
$$
= \omega_1 \wedge \alpha_1 + \omega_2 \wedge \alpha_2 - y_1
\vartheta_1 - y_2 \vartheta_2.
$$
Similarly, beginning with $\alpha_2$,
$$
d \alpha_2 =- y_1 d \omega_3 + \omega_3 \wedge d y_1 +
y_2 d \omega_1 - \omega_1 \wedge d y_2
$$
$$
=- y_1 \, \vartheta_3 + 2 y_1 \, \omega_1 \wedge \omega_3
+ \omega_3 \wedge \alpha_1 + y_1 \, \omega_3 \wedge \omega_1
+ y_2 \, \omega_3 \wedge \omega_2 + y_2 \, \vartheta_1
+ y_2  \, \omega_2 \wedge \omega_3 - \omega_1 \wedge \alpha_2
- y_1 \, \omega_1 \wedge \omega_3
$$
$$
= - \omega_1 \wedge \alpha_2 + \omega_3 \wedge \alpha_1 + y_2 \,
\vartheta_1 - y_1 \, \vartheta_3,
$$
as required.

{\bf Corollary 2.1.} The exterior derivatives of the $\alpha_i$ 
in \eqref{eq2.9} can be expressed concisely in terms of the
matrix elements of $\Theta$ and $\Omega$ as
\be
d \alpha_i =- \Theta_{ij} y_j + \Omega_{ij} \wedge \alpha_j,
\qquad
i,j=1,2.
\label{eq2.11}
\end{equation}

The Corollary is Theorem 2.1 after using the definitions of the matrices
in \eqref{eq2.2} and \eqref{eq2.4}.

The forms $\alpha_1, \alpha_2$ given in \eqref{eq2.9} lead to a natural
Riccati representation called $\alpha_3$, $\alpha_4$ for this differential
system. This arises by taking the following
linear combinations of $\alpha_1$ and $\alpha_2$,
\be
y_1^2 \, \alpha_3 = y_1 \alpha_2 - y_2 \alpha_1
= y_1 \, d y_2 - y_2 \, d y_1 - y_1^2 \, \omega_3 +2 y_1 y_2
\omega_1 + y_2^2 \omega_2,
\label{eq2.12}
\end{equation}
\be
y_2^2 \, \alpha_4 = y_2 \alpha_1 - y_1 \alpha_2 = y_2 \, d y_1
- y_1 \, d y_2 - y_2^2 \omega_2 - 2 y_1 y_2 \omega_1 + y_1^2
\omega_3.
\label{eq2.13}
\end{equation}
Two new functions, or pseudopotentials, $y_3$ and $y_4$
can be introduced which are defined by the transformation
\be
y_3 = \frac{y_2}{y_1},
\qquad
y_4 = \frac{y_1}{y_2}.
\label{eq2.14}
\end{equation}
Then, from \eqref{eq2.12} and \eqref{eq2.13}, the following
Riccati forms result,
\be
\alpha_3 = d y_3 - \omega_3 + 2 y_3 \omega_1 + y_3^2 \omega_2,
\qquad
\alpha_4 = d y_4 - \omega_2 - 2 y_4 \omega_1 + y_4^2 \omega_3.
\label{eq2.15}
\end{equation}
These equations could also be thought of as Riccati equations 
for $y_3$ and $y_4$.

{\bf Theorem 2.2.} $(i)$ The one-forms $\alpha_3$, $\alpha_4$
given by \eqref{eq2.15} have  exterior derivatives 
\be
d \alpha_3 = 2 y_3 \, \vartheta_1 + y_3^2 \, \vartheta_2 - \vartheta_3
+ 2 \alpha_3 \wedge ( \omega_1 + y_3 \, \omega_2),
\quad
d \alpha_4 =- 2 y_4 \vartheta_1 - \vartheta_2 + y_4^2 \vartheta_3
+ 2 \alpha_4 \wedge ( - \omega_1 + y_4 \omega_3).
\label{eq2.16}
\end{equation}
Consequently, results \eqref{eq2.16} are contained in the ring
spanned by $\{ \vartheta_i \}$ and $\{ \alpha_j \}$.

$(ii)$ Define the forms $\sigma_1 = \omega_1 + y_3 \omega_2$ and
$\sigma_2 =- \omega_1 + y_4 \omega_3$ appearing in \eqref{eq2.16}.
The exterior derivatives of $\sigma_1$ and $\sigma_2$ are found to be
$$
d \sigma_1 = \vartheta_1 + y_3 \vartheta_2 + \alpha_3 \wedge \omega_2,
\qquad
d \sigma_2 = - \vartheta_1 + y_3 \vartheta_2 + \alpha_4 \wedge
\omega_2.
$$
Therefore $d \sigma_1 \equiv 0$ and $d \sigma_2 \equiv 0$
$\mod \{ \vartheta_i, \alpha_j \}$.

The proof of this Theorem proceeds along exactly the same
lines as Theorem 2.1. The $S L (2, \mathbb R)$ structure and
the connection interpretation can be based on the forms
$\alpha_3$, $\alpha_4$. 

At this point, the process can be continued. Two more 
pseudopotentials $y_5$ and $y_6$ can be introduced and the
differential system can be extended by including two more 
one-forms $\alpha_5$ and $\alpha_6$. The specific structure
of these forms is suggested by Theorem 2.2.

{\bf Theorem 2.3.} Define two one-forms $\alpha_5$ and
$\alpha_6$ as
\be
\alpha_5 = d y_5 -  \omega_1 - y_3 \omega_2,
\qquad
\alpha_6 = d y_6 + \omega_1 - y_4 \omega_3.
\label{eq2.17}
\end{equation}
The exterior derivatives of the one-forms \eqref{eq2.17}
are given by the expressions
\be
d \alpha_5 =- \vartheta_1 - y_3 \vartheta_3 - \alpha_3 \wedge \omega_2,
\quad
d \alpha_6 = \vartheta_1 - y_4 \vartheta_3 - \alpha_4 \wedge \omega_3.
\label{eq2.18}
\end{equation}
Results \eqref{eq2.18} specify the closure properties of
the forms $\alpha_5$ and $\alpha_6$.
This theorem is proved along similar lines keeping in mind the results
for $d y_3$ and $d y_4$ are obtained from \eqref{eq2.15}.

A final extension can be made by adding two additional pseudopotentials
$y_7$ and $y_8$. At each stage, the ideal of forms is being
enlarged and is found to be closed over the underlying
subideal which does not contain the two new forms.

{\bf Theorem 2.4.} Define the one-forms $\alpha_7$ and
$\alpha_8$ by means of
\be
\alpha_7 = d y_7 - e^{- 2 y_5} \, \omega_2,
\qquad
\alpha_8 = d y_8 - e^{-2y_6} \, \omega_3.
\label{eq2.19}
\end{equation}
The exterior derivatives of the forms $\alpha_7$ and $\alpha_8$
are given by
\be
d \alpha_7 = 2 e^{-2 y_5} \, \alpha_5 \wedge \omega_2 - e^{-2 y_5} 
\, \vartheta_2,
\qquad
d \alpha_8 = 2 e^{-2 y_6} \alpha_6 \wedge \omega_3 - e^{-2 y_6} \,
\vartheta_3,
\label{eq2.20}
\end{equation}
and are closed over the prolonged ideal.

\begin{center}
{\bf 3. $O(3)$ Prolongations.}
\end{center}

The first $3 \times 3$ problem we consider is formulated in
terms of a set of one-forms $\omega_i$ for $i=1,2,3$.
The following system of Pfaffian equations is to be
associated to a specific nonlinear equation
\be
\alpha_i =0,
\qquad
\alpha_i = d y_i - \Omega_{ij} y_j,
\qquad
i,j=1,2,3.
\label{eq3.1}
\end{equation}
In \eqref{eq3.1}, $\Omega$ is a traceless $3 \times 3$ matrix
of one-forms. The nonlinear equation to emerge is expressed
as the vanishing of a traceless $3 \times 3$ matrix of two-forms
$\Theta$,
\be
\Theta =0,
\qquad
\Theta = d \Omega - \Omega \wedge \Omega.
\label{eq3.2}
\end{equation}
These can be considered as integrability conditions for
\eqref{eq3.1}.  The closure property, the gauge transformation 
and the gauge theoretic interpretation can also hold for
$3 \times 3$ systems. Explicitly the matrix $\Omega$ is
given by
\be
\Omega = 
\begin{pmatrix}
0  &  - \omega_1  &  \omega_2   \\
\omega_1  &  0  &  - \omega_3   \\
- \omega_2  &  \omega_3  &  0   \\
\end{pmatrix}
\label{eq3.3}
\end{equation}
Using \eqref{eq3.3} in \eqref{eq3.1} and \eqref{eq3.2},
the following prolonged differential system is obtained,
\be
\begin{array}{ccc}
d y_1 = \alpha_1 -y_2 \omega_1 + y_3 \omega_2,  &
&  d \omega_1 = \vartheta_1 - \omega_2 \wedge \omega_3,  \\
  &  &   \\
d y_2 = \alpha_2 + y_1 \omega_1 - y_3 \omega_3, &
&  d \omega_2 = \vartheta_2 - \omega_3 \wedge \omega_1,  \\
  &  &   \\
d y_3 = \alpha_3 - y_1 \omega_2 + y_2 \omega_3, &
& d \omega_3 = \vartheta_3 - \omega_1 \wedge \omega_2.   \\
\end{array}
\label{eq3.4}
\end{equation}
By straightforward exterior differentiation of each
$\alpha_i$ and using \eqref{eq3.4}, the following
Theorem results.

{\bf Theorem 3.1.} The $\alpha_i$ given in \eqref{eq3.4}
have exterior derivatives which are contained in the
ring spanned by $\{ \vartheta_i \}_1^3$ and 
$\{ \alpha_j \}_1^3$ and given explicitly as
\be
\begin{array}{c}
d \alpha_1 = y_2 \vartheta_1 - y_3 \vartheta_2 
- \omega_1 \wedge \alpha_2 + \omega_2 \wedge \alpha_3,  \\
 \\
d \alpha_2 =- y_1 \vartheta_1 + y_3 \vartheta_3
+ \omega_1 \wedge \alpha_1 - \omega_3 \wedge \alpha_3,  \\
 \\
d \alpha_3 = y_1 \vartheta_2 - y_2 \vartheta_3
- \omega_2 \wedge \alpha_1 + \omega_3 \wedge \alpha_2.
\end{array}
\label{eq3.5}
\end{equation}

The one-forms $\{ \alpha_i \}_1^3$ admit a series
of Riccati representations which can be realized by 
defining, six new one-forms,
$$
y_1^2 \alpha_4 = y_1 \alpha_2 - y_2 \alpha_1,
\quad
y_1^2 \alpha_5 = y_1 \alpha_3 - y_3 \alpha_1,
\quad
y_2^2 \alpha_6 = y_2 \alpha_1 - y_1 \alpha_2,
$$
$$
y_2^2 \alpha_7 = y_2 \alpha_3 - y_3 \alpha_2,
\quad
y_3^2 \alpha_8 = y_3 \alpha_1 - y_1 \alpha_3,
\quad
y_3^2 \alpha_9 = y_3 \alpha_2 - y_2 \alpha_3.
$$
In effect, the larger $3 \times 3$ system is breaking up
into several $2 \times 2$ systems. 
To write the new one-forms explicitly, introduce the
new functions $\{ y_j \}_4^9$ which are defined in terms
of the original $\{ y_i \}_1^3$ as follows,
\be
y_4 = \frac{y_2}{y_1},
\quad
y_5 = \frac{y_3}{y_1},
\quad
y_6 = \frac{y_1}{y_2},
\quad
y_7 = \frac{y_3}{y_1},
\quad
y_8 = \frac{y_1}{y_3},
\quad
y_9 = \frac{y_2}{y_3}.
\label{eq3.6}
\end{equation}
In terms of the functions defined in \eqref{eq3.6}, the
$\{ \alpha_j \}_4^9$ are given as,
\be
\begin{array}{cc}
\alpha_4 = d y_4 - ( 1 + y_4^2) \omega_1 + y_4 y_5
\omega_2 + y_5 \omega_3,
&
\alpha_5 = d y_5 - y_4 y_5 \omega_1 + ( 1 + y_5^2) \omega_2
- y_4 \omega_3,     \\
   &         \\
\alpha_6 = d y_6 + (1 + y_6^2) \omega_1 - y_7 \omega_2
- y_6 y_7 \omega_3,
&
\alpha_7 = d y_7 + y_6 y_7 \omega_1 + y_6 \omega_2
- ( 1 + y_7^2) \omega_3,   \\    
   &      \\
\alpha_8 = d y_8 + y_9 \omega_1 - ( 1 + y_8^2) \omega_2
+ y_8 y_9 \omega_3,
&
\alpha_9 = d y_9 - y_8 \omega_1 - y_8 y_9 \omega_2
+ ( 1 + y_9^2) \omega_3.  \\
\end{array}
\label{eq3.7}
\end{equation}
To state the next theorem, we need to define the following
three matrices,
\be
\Omega_1 =
\begin{pmatrix}
2 y_4 \omega_1 - y_5 \omega_2  &  - y_4 \omega_2 - \omega_3  \\
y_5 \omega_1 + \omega_3  & y_4 \omega_1 - 2 y_5 \omega_2  \\
\end{pmatrix},
\quad
\Omega_2 =
\begin{pmatrix}
-2 y_6 \omega_1 + y_7 \omega_3  &  \omega_2 + y_6 \omega_3  \\
- y_7 \omega_1 - \omega_6  & - y_6 \omega_1 + 2 y_7 \omega_3  \\
\end{pmatrix},
\label{eq3.8}
\end{equation}
$$
\Omega_3 =
\begin{pmatrix}
2 y_8 \omega_2 - y_9 \omega_3  &  - \omega_1 - y_9 \omega_3  \\
\omega_1 + y_9 \omega_2  &  y_8 \omega_2 - 2 y_9 \omega_3 \\
\end{pmatrix}.
$$

{\bf Theorem 3.2.} The closure properties for the forms
$\{ \alpha_i \}_4^9$ given in \eqref{eq3.7} can be summarized in the
form
$$
d \begin{pmatrix}
\alpha_4  \\
\alpha_5   \\
\end{pmatrix} 
= \begin{pmatrix}
- ( 1 + y_4^2) \vartheta_1 + y_4 y_5 \vartheta_2 + y_5 \vartheta_3  \\
- y_4 y_5 \vartheta_1 + ( 1 + y_5^2) \vartheta_2 - y_4 \vartheta_3  \\
\end{pmatrix}
+ \Omega_1 \wedge \begin{pmatrix}
\alpha_4  \\
\alpha_5  \\
\end{pmatrix},
$$
\be
d \begin{pmatrix}
\alpha_6  \\
\alpha_7   \\
\end{pmatrix}
= \begin{pmatrix}
( 1 + y_6^2) \vartheta_1 - y_7 \vartheta_2 - y_6 y_7 \vartheta_3  \\
y_6 y_7 \vartheta_1 + y_6 \vartheta_2 - (1+ y_7^2) \vartheta_3  \\
\end{pmatrix}
+ \Omega_2 \wedge \begin{pmatrix}
\alpha_6  \\
\alpha_7  \\
\end{pmatrix},
\label{eq3.9}
\end{equation}
$$
d \begin{pmatrix}
\alpha_8   \\
\alpha_9   \\
\end{pmatrix}
= \begin{pmatrix}
y_1 \vartheta_1 - ( 1 +y_8^2) \vartheta_2 + y_8 y_9 \vartheta_3  \\
- y_8 \vartheta_1 - y_8 y_9 \vartheta_2 + ( 1 + y_9^2) \vartheta_3 \\
\end{pmatrix} + \Omega_3 \wedge
\begin{pmatrix}
\alpha_8  \\
\alpha_9  \\
\end{pmatrix}.
$$
The exterior derivatives are therefore contained in the ring
spanned by $\{ \vartheta_i \}_1^3$ and the $\{ \alpha_j \}_4^9$.

In terms of $\Omega_i$, a corresponding matrix of two-forms
$\Theta_i$ can be defined in terms of the corresponding
$\Omega_i$ defined in \eqref{eq3.8},
\be
\Theta_i = d \Omega_i - \Omega_i \wedge \Omega_i, \qquad i=1,2,3.
\label{eq3.10}
\end{equation}

{\bf Theorem 3.3.} The $2 \times 2$ matrix of two-forms
$\Theta_i$ defined by \eqref{eq3.10} 
is contained in the ring of forms spanned by
$\{ \vartheta_i \}_1^3$ coupled with either $\{ \alpha_4, \alpha_5 \}$,
$\{ \alpha_6, \alpha_7 \}$ or $\{ \alpha_8, \alpha_9 \}$
when $i=1,2,3$, respectively.

{\bf Proof:} Differentiating $\Omega_1$ and simplifying, we have
$$
\Theta_1 = d \Omega_1 - \Omega_1 \wedge \Omega_1
$$
$$
= \begin{pmatrix}
2 y_4 \vartheta_1 - y_5 \vartheta_2 + 2 \alpha_4 \wedge \omega_1 - \alpha_5 \wedge \omega_2  
&  -y_4 \vartheta_2 - \vartheta_3 - \alpha_4 \wedge \omega_2  \\
y_5 \vartheta_1 + \vartheta_3 + \alpha_5 \wedge \omega_1  &
y_4 \vartheta_1 - 2 y_5 \vartheta_2 + \alpha_4 \wedge \omega_1 - 2 \alpha_5 \wedge \omega_2  \\
\end{pmatrix}.
$$
Similar matrix expressions can be found for the cases 
in which $\Omega_1$ is replaced by $\Omega_2$ or $\Omega_3$,
respectively.

Unlike $\Omega$ given in \eqref{eq3.3}, the $\Omega_i$ in
\eqref{eq3.8} are not traceless. Define their traces to be
\be
\kappa_i = tr \, \Omega_i.
\qquad
i=1,2,3.
\label{eq3.11}
\end{equation}
These traces provide convenient ways of generating conservation laws
on account of the following Theorem.

{\bf Theorem 3.4.} The exterior derivatives of traces \eqref{eq3.11}
are given by
$$
\frac{1}{3} d \kappa_1 = y_4 \vartheta_1 - y_5 \vartheta_2 
- \omega_1 \wedge \alpha_4 + \omega_2 \wedge \alpha_5,
\quad
\frac{1}{3} d \kappa_2 =- y_6 \vartheta_1 + y_7 \vartheta_3
+ \omega_1 \wedge \alpha_6 - \omega_3 \wedge \alpha_7,
$$
\be
\frac{1}{3} d \kappa_3 =y_8 \vartheta_2 - y_9 \vartheta_3
- \omega_2 \wedge \alpha_8 + \omega_3 \wedge \alpha_9.
\label{eq3.12}
\end{equation}
Results \eqref{eq3.12} are contained in the ring
spanned by $\{ \vartheta_i \}_1^3$ and $\{ \alpha_j \}_4^9$.

\begin{center}
{\bf 4. $SU (3)$ Prolongations.}
\end{center}

To formulate a $3 \times 3$ $SU (3)$ problem, the set of Pfaffian equations
\be
\alpha_i =0,
\qquad
\alpha_i = d y_i - \Omega_{ij} y_j,
\qquad
i,j=1,2,3,
\label{eq4.1}
\end{equation}
are associated to the nonlinear equation. In \eqref{eq4.1}, $\Omega$ 
is a traceless $3 \times 3$ matrix consisting of a system of
one-forms. The nonlinear equation to be considered is expressed
as the vanishing of a traceless $3 \times 3$ matrix of two-forms
$\Theta$ exactly as in \eqref{eq3.2} which constitute the
integrability condition for \eqref{eq4.1}.

A $3 \times 3$ matrix representation of the Lie algebra for
$SU (3)$ is introduced by means of generators
$\lambda_j$ for $j=1, \cdots, 8$,
which satisfy the following set of commutation relations
\be
[ \lambda_l, \lambda_m ] = 2i \, f_{lmn} \lambda_n.
\label{eq4.2}
\end{equation}
The structure constants $f_{lmn}$ are totally anti-symmetric
in $l,m,n$. The one-form $\Omega$ is expressed in terms of the
$\lambda_i$ as
\be
\Omega = \sum_{l=1}^8 \, \omega_l \, \lambda_l.
\label{eq4.3}
\end{equation}
Then the two-form $\Theta$ is written in the form
\be
\Theta = \sum_{l=1}^8 \, \vartheta_l \lambda_l,
\qquad
\vartheta_l = d \omega_l -i f_{lmn} \, \omega_m \wedge \omega_n.
\label{eq4.4}
\end{equation}
The nonlinear equation to be solved has the form
$\vartheta_l =0$ for $l=1, \cdots, 8$.

It will be useful to display \eqref{eq4.3} and \eqref{eq4.4}
by using the nonzero structure constants given in \eqref{eq4.2},
since these may not be readily accessible. An explicit 
representation for the matrix $\Omega$ used here is given by
\be
\Omega = \begin{pmatrix}
\omega_3 + \frac{1}{\sqrt{3}} \omega_8 & \omega_1 -i \omega_2 &
\omega_4 -i \omega_5  \\
\omega_1 +i \omega_2 & - \omega_3 + \frac{1}{\sqrt{3}} \omega_8  & 
\omega_6 -i \omega_7   \\
\omega_4 +i \omega_5  & \omega_6 +i \omega_7  & - \frac{2}{\sqrt{3}} \omega_8  \\
\end{pmatrix}
\label{eq4.5}
\end{equation}
Moreover, in order that the presentation be easier to follow,
the eight forms $d \omega_i$ specified by \eqref{eq4.4} will be given explicitly,
\be
\begin{array}{c}
d \omega_1 = \vartheta_1 +2 i \omega_2 \wedge \omega_3 +i \omega_4 \wedge
\omega_7 -i \omega_5 \wedge \omega_6, \\ 
 \\
d \omega_2 = \vartheta_2 - 2 i \omega_1 \wedge \omega_3 + i
\omega_4 \wedge \omega_6 + i \omega_5 \wedge \omega_7,   \\
  \\
d \omega_3 = \vartheta_3 + 2 i \omega_1 \wedge \omega_2 +
i \omega_4 \wedge \omega_5 -i \omega_6 \wedge \omega_7,  \\
  \\
d \omega_4 = \vartheta_4 -i \omega_1 \wedge \omega_7 -i \omega_2
\wedge \omega_6 -i \omega_3 \wedge \omega_5 + \sqrt{3} i
\omega_5 \wedge \omega_8,  \\
  \\
d \omega_5 = \vartheta_5 + i \omega_1 \wedge \omega_6 - i \omega_2 \wedge 
\omega_7 +i \omega_3 \wedge \omega_4 - \sqrt{3} i \omega_4 \wedge \omega_8,  \\
  \\
d \omega_6 = \vartheta_6 -i \omega_1 \wedge \omega_5 + i \omega_2 \wedge \omega_4 
+i \omega_3 \wedge \omega_7 + \sqrt{3} i \omega_7 \wedge \omega_8,  \\
  \\
d \omega_7 = \vartheta_7 + i \omega_1 \wedge \omega_4 + i \omega_2 
\wedge \omega_5 -i \omega_3 \wedge \omega_6 - \sqrt{3} i \omega_6 
\wedge \omega_8,   \\
  \\
d \omega_8 = \vartheta_8 + \sqrt{3} i \omega_4 \wedge \omega_5 +
\sqrt{3} i \omega_6 \wedge \omega_7.
\end{array}
\label{eq4.6}
\end{equation}
Substituting \eqref{eq4.5} into \eqref{eq4.1}, the following
system of one-forms $ \{ \alpha_i \}_1^3$ is obtained,
\be
\begin{array}{c}
\alpha_1 = d y_1 - ( \omega_3 + \frac{1}{\sqrt{3}} \omega_8) y_1
- (\omega_1 -i \omega_2 ) y_2 - ( \omega_4 -i \omega_5) y_3, \\
   \\
\alpha_2 = d y_2 - (\omega_1 +i \omega_2) y_1 
+( \omega_3 - \frac{1}{\sqrt{3}} \omega_8) y_2 - (\omega_6 -i \omega_7) y_3,  \\
   \\
\alpha_3 = d y_3 - ( \omega_4 + i \omega_5 ) y_1 - (\omega_6 +i \omega_7 ) y_2
+ \dss \frac{2}{\sqrt{3}} \omega_8 y_3.
\end{array}
\label{eq4.7}
\end{equation}
The increase in complexity of this system makes it often
necessay to resort to the use of symbolic manipulation
to carry out longer calculations, and for the most part, only
results are given..

{\bf Theorem 4.1.} The exterior derivatives of the $\alpha_i$
in \eqref{eq4.7} are given by
$$
d \alpha_1 = - y_2 (\vartheta_1 -i \vartheta_2) - y_1 \vartheta_3
- y_3 ( \vartheta_4 -i \vartheta_5) - \frac{1}{\sqrt{3}} y_1 \vartheta_8
+ ( \omega_3 + \frac{1}{\sqrt{3}} \omega_8) \wedge \alpha_1
+ ( \omega_1 -i \omega_2) \wedge \alpha_2 + ( \omega_4 -i \omega_5)
\wedge \alpha_3,
$$
\be
d \alpha_2 =- y_1 ( \vartheta_1 +i \vartheta_2) + y_2 \vartheta_3
- \frac{1}{\sqrt{3}} y_2 \vartheta_8 - y_3 ( \vartheta_6 -i \vartheta_7)
+(\omega_1 +i \omega_2) \wedge \alpha_1 - ( \omega_3 - \frac{1}{\sqrt{3}}
\omega_8) \wedge \alpha_2 + ( \omega_6 -i \omega_7) \wedge \alpha_3,
\label{eq4.8}
\end{equation}
$$
d \alpha_3 =- y_1 ( \vartheta_4 + i \vartheta_5) - y_2 ( \vartheta_6 +i 
\vartheta_7) + \frac{2}{\sqrt{3}} y_3 \vartheta_8 + ( \omega_4 +i
\omega_5) \wedge \alpha_1 + ( \omega_6 +i \omega_7) \wedge \alpha_2
- \frac{2}{\sqrt{3}} \omega_8 \wedge \alpha_3.
$$

In order to keep the notation concise, an abbreviation
will be introduced. Suppose $\{ \gamma_i \}$ is a system of forms,
and $i, j$ are integers, then we make abbreviation
$$
\gamma_{ij \pm} = \gamma_i \pm i \gamma_j.
$$
At this point, quadratic pseudopotentials can be introduced in
terms of the homogeneous variables of the same form as \eqref{eq3.6}. In the same way
that \eqref{eq3.7} was produced, the following Pfaffian equations
based on the set $\{ \alpha_i \}_1^3$ in \eqref{eq4.7} are obtained,
\be
\begin{array}{c}
\alpha_4 = d y_4 - \omega_{12+} + 2 y_4 \omega_3 + y_4^2 \omega_{12-}
- y_5 \omega_{67-} + y_4 y_5 \omega_{45-},  \\
  \\
\alpha_5 = d y_5 - \omega_{45+} + y_5 ( \omega_3 + \sqrt{3} \omega_8)
+ y_5^2 \omega_{45-} - y_4 \omega_{67+} + y_4 y_5 \omega_{12-},   \\
  \\
\alpha_6 = d y_6 - \omega_{12-} - 2 y_6 \omega_3 + y_6^2 \omega_{12+}
- y_7 \omega_{45-} + y_6 y_7 \omega_{67-},   \\
  \\
\alpha_7 = d y_7 - \omega_{67+} - y_7 (\omega_3 - \sqrt{3} \omega_8)
+ y_7^2 \omega_{67-} - y_6 \omega_{45+} + y_6 y_7 \omega_{12+},  \\
  \\
\alpha_8 = d y_8 - \omega_{45-} - y_8 (\omega_3 + \sqrt{3} \omega_8)
+ y_8^2 \omega_{45+} - y_9 \omega_{12-} + y_8 y_9 \omega_{67+},  \\
  \\
\alpha_9 = d y_9 - \omega_{67-} +y_9  ( \omega_3 - \sqrt{3} \omega_8)
+y_9^2 \omega_{67+} -   y_8 \omega_{12+} +y_8 y_9 \omega_{45+}.
\end{array}
\label{eq4.9}
\end{equation}
These are coupled Riccati equations for the pairs of pseudopotentials
$(y_4, y_5)$, $(y_6, y_7)$ and $(y_8, y_9)$, respectively.

Define the following set of $2 \times 2$ matrices of one-forms
$\Omega_i$ for $i=1,2,3$ as
$$
\Omega_1 = \begin{pmatrix}
-2 \omega_3 -2 y_4 \omega_{12-} - y_5 \omega_{45-} &
\omega_{67-} - y_4 \omega_{45-}  \\
\omega_{67+} - y_5 \omega_{12-} &
- \omega_3 - \sqrt{3} \omega_8 - y_4 \omega_{12-}
- 2 y_5 \omega_{45-}   \\
\end{pmatrix}
$$
\be
\Omega_2 = \begin{pmatrix}
2 \omega_3 -2 y_6 \omega_{12+} - y_7 \omega_{67-} &
\omega_{45-} - y_6 \omega_{67-}    \\
\omega_{45+} - y_7 \omega_{12 +}  &
\omega_3 - \sqrt{3} \omega_8 - y_6 \omega_{12+} -2 y_7
\omega_{67-}   \\
\end{pmatrix}
\label{eq4.10}
\end{equation}
$$
\Omega_3 = \begin{pmatrix}
\omega_3 + \sqrt{3} \omega_8 - 2 y_8 \omega_{45+} - y_9 \omega_{67+} &
\omega_{12-} - y_8 \omega_{67+}  \\
\omega_{12+} - y_9 \omega_{45+}  &
- \omega_3 + \sqrt{3} \omega_8 - y_8 \omega_{45+} -2  y_9 \omega_{67+} \\
\end{pmatrix}
$$
Making use of the matrices $\Omega_i$ defined in \eqref{eq4.10}, the
following result can be stated.

{\bf Theorem 4.2.} The closure properties of the set of forms
$\{ \alpha_i \}_4^9$ given by \eqref{eq4.9} can be expressed in terms 
of the $\Omega_i$ as follows,
$$
d \begin{pmatrix}
\alpha_4  \\
\alpha_5  \\
\end{pmatrix} = 
\begin{pmatrix}
(y_4^2 -1) \vartheta_1 -i ( y_4^2 +1) \vartheta_2 + 2 y_4 \vartheta_3
+ y_4 y_5 \vartheta_{45-} - y_5 \vartheta_{67-}  \\
y_4 y_5 \vartheta_{12-} + y_5 \vartheta_3 + ( y_4^2 -1) \vartheta_4
- i (y_5^2 +1) \vartheta_5 - y_4 \vartheta_{67+} + \sqrt{3} y_5 \vartheta_8 \\
\end{pmatrix}
+ \Omega_1 \wedge \begin{pmatrix}
\alpha_4  \\
\alpha_5   \\
\end{pmatrix},
$$
\be 
d \begin{pmatrix}
\alpha_6   \\
\alpha_7   \\
\end{pmatrix} =
\begin{pmatrix}
(y_6^2 -1) \vartheta_1 +i (y_6^2+1) \vartheta_2 -2 y_6 \vartheta_3
-y_7 \vartheta_{45-} + y_6 y_7 \vartheta_{67-}  \\
y_6 y_7 \vartheta_{12+} - y_7 \vartheta_3 - y_6 \vartheta_{45+}
+ ( y_7^2 -1) \vartheta_6 -i (y_7^2+1) \vartheta_7 +\sqrt{3} y_7 
\vartheta_8   \\
\end{pmatrix} + \Omega_2 \wedge 
\begin{pmatrix}
\alpha_6  \\
\alpha_7  \\
\end{pmatrix}
\label{eq4.11}
\end{equation}
$$
d \begin{pmatrix}
\alpha_8   \\
\alpha_9   \\
\end{pmatrix} =
\begin{pmatrix}
- y_9 \vartheta_{12-} -y_8 \vartheta_3 + ( y_8^2 -1) \vartheta_4
+i (y_8^2+1) \vartheta_5 + y_8 y_9
\vartheta_{67+} - \sqrt{3} y_8 \vartheta_8  \\
- y_8 \vartheta_{12+} + y_9 \vartheta_3 + y_8 y_9 \vartheta_{45+}
+ (y_9^2-1) \vartheta_6 + i (y_9^2 +1) \vartheta_7 - \sqrt{3} y_9
\vartheta_8 \\
\end{pmatrix}
+ \Omega_3 \wedge \begin{pmatrix}
\alpha_8  \\
\alpha_9  \\
\end{pmatrix}.
$$

Based on the forms $\Omega_i$ given in \eqref{eq4.10}, we can
differentiate to define the following matrices of two-forms $\xi_i$ given by
\be
\xi_i = d \Omega_i - \Omega_i \wedge \Omega_i.
\label{eq4.12}
\end{equation}
An analogue of Theorem 3.3 can now be formulated.

{\bf Theorem 4.3.} The forms $\xi_i$ defined in \eqref{eq4.12}
are given explicitly by the matrices
$$
\xi_i = \begin{pmatrix}
2 \omega_{12+} \wedge \alpha_4 + \omega_{45-} \wedge \alpha_5
- 2 y_4 \vartheta_{12-}  &
\omega_{45-} \wedge \alpha_4 - y_4 \vartheta_{45-} + \vartheta_{67-}  \\
-2 \vartheta_3 - y_5 \vartheta_{45-}  &              \\
  &  \omega_{12-} \wedge \alpha_4 + 2 \omega_{45-} \wedge \alpha_5
- y_4 \vartheta_{12-} - \vartheta_3   \\
\omega_{12-} \wedge \alpha_5 - y_5 \vartheta_{12-} + \vartheta_{67+}
         &  - 2 \vartheta_{45-} - \sqrt{3} \vartheta_8 \\
\end{pmatrix}
$$
\be
\xi_2 =
\begin{pmatrix}
2 \omega_{12+} \wedge \alpha_6 + \omega_{67-} \wedge \alpha_7 - 2 y_6 \vartheta_{12+}  &
\omega_{67-} \wedge \alpha_6 + \vartheta_{45-} - y_6 \vartheta_{67-}   \\
+ \vartheta_3 - y_7 \vartheta_{67-}    &       \\
 &\omega_{12+} \wedge \alpha_6 + 2 \omega_{67-} \wedge \alpha_7 - y_6 \vartheta_{12+} + \vartheta_3  \\
\omega_{12+} \wedge \alpha_7 - y_7 \, \vartheta_{12+} + \vartheta_{45+}     
&  - 2 y_7 \vartheta_{67-} - \sqrt{3} \vartheta_8  \\
\end{pmatrix}
\label{eq4.13}
\end{equation}
$$
\xi_3 = \begin{pmatrix}
2 \omega_{45+} \wedge \alpha_8 + \omega_{67+} \wedge \alpha_9 + \vartheta_3 - 2 y_8 \vartheta_{45+} 
& \omega_{67+} \wedge \alpha_8 + \vartheta_{12-} - y_8 \vartheta_{67+}  \\
- y_9 \vartheta_{67+} + \sqrt{3} \vartheta_8   &    \\
 & \omega_{45+} \wedge \alpha_8 + 2 \omega_{67+} \wedge \alpha_9 - \vartheta_3 - y_8 \vartheta_{45+}  \\
\omega_{45+} \wedge \alpha_9 + \vartheta_{12+} - y_9 \vartheta_{45+}   
&   - 2 y_9 \vartheta_{67+} + \sqrt{3} \vartheta_8   \\
\end{pmatrix}
$$
From these results, it is concluded that $\xi_1$ is contained in the
ring of $\{ \alpha_4, \alpha_5 \}$ and the $\{ \vartheta_l \}$, $\xi_2$
is contained in the ring of $\{ \alpha_6, \alpha_7 \}$ and $\{ \vartheta_l \}$,
and $\xi_3$ is in the ring of $\{ \alpha_8, \alpha_9 \}$ and $\{ \vartheta_l \}$.

The two-forms $\{ \xi_i \}_1^3$ therefore vanish for the solutions of the 
nonlinear equations \eqref{eq4.2} and of the coupled Riccati
equations $\{ \alpha_j =0 \}_4^9$. As in the case of $O (3)$, the
one-forms $\Omega_i$ are not traceless. Denoting the trace of
$\Omega_i$ by
\be
\tau_i = tr \, \Omega_i,
\label{eq4.14}
\end{equation}
the following theorem then follows.

{\bf Theorem 4.4.} The closure properties of the exterior derivatives
of the traces $\tau_i$ are given by,
$$
\frac{1}{3} d \tau_1  =\omega_{12-} \wedge \alpha_4 + \omega_{45-} \wedge 
\alpha_5 - y_4 \vartheta_{12-} - \vartheta_3 - y_5 \vartheta_{45-}
- \frac{1}{\sqrt{3}} \vartheta_8,
$$
\be
\frac{1}{3} d \tau_2 = \omega_{12+} \wedge \alpha_6 + \omega_{67-}
\wedge \alpha_7 - y_6 \vartheta_{12+} + \vartheta_3 - y_7
\vartheta_{67-} - \frac{1}{\sqrt{3}} \vartheta_8,
\label{eq4.15}
\end{equation}
$$
\frac{1}{3} d \tau_3 = \omega_{45+} \wedge \alpha_8 + \omega_{67+}
\wedge \alpha_9 - y_8 \vartheta_{45+} - y_9 \vartheta_{67+}
+ \frac{2}{\sqrt{3}} \vartheta_8.
$$

The one-forms \eqref{eq4.14} then can be used to generate 
conservation laws. Note that as in the $O(3)$ case, a
$3 \times 3$ problem $\Omega$ has been reduced to three
separate $2 \times 2$ problems in terms of the matrices $\Omega_l$.
Further prolongation can be continued beginning with the
one-forms $\Omega_l$ given in \eqref{eq4.10}. To briefly outline
the further steps in the procedure, a system of 
Pfaffians $\{ \tilde{\alpha}_{lj} \}$ 
are introduced which are based on the one-forms $\Omega_l$
\be
\tilde{\alpha}_{lj} =0,
\qquad
\tilde{\alpha}_{lj} = d y_{lj} - ( \Omega_l )_{jk} y_{lk}.
\label{eq4.16}
\end{equation}
The subscript $l$ in $y_{lj}$ and $\tilde{\alpha}_{lj}$ indicates
that they belong to the sub-system defined by $\Omega_l$, 
and it is not summed over. Exterior differentiation of $\tilde{\alpha}_{lj}$
yields 
\be
d \tilde{\alpha}_{lj} = (\Omega_l)_{jk} \wedge \tilde{\alpha}_{lk}
- ( \tilde{\Theta} )_{jk} y_{lk}.
\label{eq4.17}
\end{equation}
The $\{ y_{lj} \}$ are pseudopotentials for the original 
nonlinear equation \eqref{eq4.2}. Next, quadratic 
pseudopotentials are included as the homogeneous variables
and the prolongation can be continued.

\begin{center}
{\bf 5. A Conservation Law and Summary.}
\end{center}

These types of results turn out to be very useful 
for further study of integrable equations.
They can be used for generating infinite numbers of conservation laws. 
In addition, the results are independent of any further structure of
the forms $\omega_i$
in all cases. In fact, the one-form $\Omega$ need not be 
thought of as unique. This is due to the fact that
$\Omega$ and $\Theta$ are form invariant under the
gauge transformation $\Omega \raro \omega' = d A \, A^{-1}
+ A \Omega A^{-1}$ and $\Theta \raro \Theta' = A \Theta A^{-1}$
where, in the $SL (2, \mathbb R)$ case, $A$ is an arbitrary $2 \times 2$ 
space-time dependent matrix of determinant one. The gauge transformation
property holds in the $3 \times 3$ case as well.
The pseudopotentials serve as potentials for conservation laws
in a generalized sense. They can be defined under a choice
of Pfaffian forms such as
\be
\alpha_i =0,
\qquad
\alpha_i = d y_i + F_i \, dx + G_i \, dt,
\label{eq5.1}
\end{equation}
with the property that the exterior derivatives $d \alpha_i$
are contained in the ring spanned by $\{ \alpha_i \}$ and $\{ \vartheta_l \}$
\be
d \alpha_i = \sum_j \, A_{ij} \wedge \alpha_j +
\sum_{l}  \, \Gamma_{il} \vartheta_l.
\label{eq5.2}
\end{equation}
This can be thought of as a generalization of the Frobenius Theorem
for complete integrability of Pfaffian systems.

Theorem 2.2 implies that
$d \sigma_i =0$ when $i=1,2$ for solutions of the original equation.
If either of these forms 
is expressed as $\sigma = {\cal I} \, dx + {\cal J} \, dt$, then
$ d \sigma=0$ imlies that
\be
\frac{\partial {\cal I}}{\partial t}
- \frac{\partial {\cal J}}{\partial t} =0.
\label{eq5.3}
\end{equation}
Thus ${\cal I}$ is a conserved density and ${\cal J}$ a conserved
current.
Let us finally represent $\omega_1 = a_1 \, dx + b_1 \, dt$,
$\omega_2 = a_2 \, dx + b_2 \, dt$ and $\omega_3 = \eta \, dx
+ b_3 \, dt$, where $\eta$ is a parameter. The $x$-dependent  
piece of $\alpha_3$ in \eqref{eq2.15} implies
\be
y_{3,x} + 2 a_1 y_3 + a_2 y_3^2 - \eta =0.
\label{eq5.4}
\end{equation}
By substituting an asymptotic expansion around $\eta = \infty$
for $y_3$ in \eqref{eq5.4} of the form,
$$
y_3 = \sum_{0}^{\infty}  \, \eta^{-n} Y_n,
$$
a recursion for the $Y_n$ is obtained. Thus \eqref{eq5.4} becomes,
$$
\sum_n \, \eta^{-n} Y_{n,x} + 2 a_1 \sum_{n} \, \eta^{-n} Y_n
+ a_2 ( \sum_n \, \eta^{-n} Y_n )^2 - \eta =0.
$$
Expanding the power, collecting coefficients of $\eta$ and equating
coefficients of $\eta$ to zero,  we get,
$$
Y_{1,x} + 2 a_1 Y_1 =1,
\qquad
Y_{n,x} + 2 a_1 Y_n + a_2 \sum_{k=1}^{n-1} \, Y_{n-k} Y_k =0.
$$
The consistency of solving $\alpha_3$ by using just the $x$-part is
guaranteed by complete integrability. The $x$-part of the
form $\sigma_1$ is expressed as
$$
( \sigma_1)_x = ( a_1 + \sum_{n=1}^{\infty} \, \eta^{-n} Y_n 
a_2 ) \, dx.
$$
Consequently, from \eqref{eq5.3} the $n$-th conserved density is
$$
{\cal I}_n = a_2 Y_n.
$$

The existence of links between the types of prolongation here
and other types of prolongations which are based on closed differential
systems is a subject for further work.

\begin{center}
{\bf References.}
\end{center}

\noindent
$[1]$ F. B. Estabrook, H. D. Wahlquist, Classical geometries defined by exterior
differential systems on higher frame bundles, Class. Quant. Grav. {\bf 6} (1989), 
263-274.  \\
$[2]$ H. Cartan, Differential Forms, Dover, Mineola, NY, (2006).  \\
$[3]$ M. J. Ablowitz, D. K. Kaup, A. C. Newell, H. Segur, Method for Solving the
sine-Gordon Equation, Phys. Rev. Lett. {\bf 30}, (1973), 1262-1264.  \\
$[4]$ M. J. Ablowitz, D. K. Kaup, A. C. Newell, H. Segur, Nonlinear Equations of Physical Interest,
Phys. Rev. Letts. {\bf31}, (1973), 125-127.  \\
$[5]$ R. Sasaki, Pseudopotentials for the general AKNS system, Phys. Lett. {\bf A 73},
(1979), 77-80.  \\
$[6]$ R. Sasaki, Geometric approach to soliton equations, Proc. R. Soc. Lond.,
{\bf A 373}, (1980), 373-384.  \\
$[7]$ M. Crampin, Solitons and $SL (2, \mathbb R)$, Phys. Lett. {\bf A 66}, 
(1978), 170-172.   \\
$[8]$ P. Bracken, Intrinsic formulation of geometric integrability and 
associated Riccati system generating conservation laws, Int. J. Geom.
Methods Mod. Phys. {\bf 6,5}, (2009), 825-837.  \\
$[9]$ H. D. Wahlquist, F. B. Estabrook, Prolongation structures of nonlinear 
evolution equations, J. Math. Phys. {\bf 16}, (1975), 1-7.  \\

\end{document}